\documentclass[prb, twocolumn,amsmath,amssymb,superscriptaddress]{revtex4}

\usepackage{graphicx}
\usepackage{dcolumn}
\usepackage{bm}

\begin{document}


\title{
Dynamic nuclear polarization
induced by breakdown of fractional quantum Hall effect
}

\author{M. Kawamura}
	\altaffiliation[Present address: ]
	{Advanced Science Institute, RIKEN, 2-1 Wako, Saitama 351-0198, Japan}
	\email{minoru@riken.jp}
	\affiliation{Institute of Industrial Science,
		University of Tokyo, 4-6-1 Komaba, Meguro-ku, Tokyo 153-8505, Japan}
	\affiliation{PRESTO, Japan Science and Technology Agency, 
		4-1-8 Kawaguchi, Saitama 333-0012, Japan}
\author{M. Ono}
	\affiliation{Institute of Industrial Science,
		University of Tokyo, 4-6-1 Komaba, Meguro-ku, Tokyo 153-8505, Japan}
\author{Y. Hashimoto}
	\affiliation{Institute for Solid State Physics, University of Tokyo, \\
		5-1-5 Kashiwanoha, Kashiwa 277-8581, Japan}
\author{S. Katsumoto}
	\affiliation{Institute for Solid State Physics, University of Tokyo, \\
		5-1-5 Kashiwanoha, Kashiwa 277-8581, Japan}
	\affiliation{Institute for Nano Quantum Information Electronics,
		University of Tokyo, 4-6-1 Komaba, Meguro-ku, Tokyo 153-8505, Japan}
\author{K. Hamaya}
	\affiliation{Institute of Industrial Science,
		University of Tokyo, 4-6-1 Komaba, Meguro-ku, Tokyo 153-8505, Japan}
	\affiliation{Institute for Nano Quantum Information Electronics,
		University of Tokyo, 4-6-1 Komaba, Meguro-ku, Tokyo 153-8505, Japan}
\author{T. Machida}\email{tmachida@iis.u-tokyo.ac.jp}
	\affiliation{Institute of Industrial Science,
		University of Tokyo, 4-6-1 Komaba, Meguro-ku, Tokyo 153-8505, Japan}
	\affiliation{Institute for Nano Quantum Information Electronics,
		University of Tokyo, 4-6-1 Komaba, Meguro-ku, Tokyo 153-8505, Japan}
\date{\today}

\begin{abstract}
We study dynamic nuclear polarization (DNP) induced by breakdown
of the fractional quantum  Hall (FQH) effect.
We find that voltage-current characteristics
depend on current sweep rates
at the quantum Hall states of Landau level filling factors $\nu$ = 1, 2/3, and 1/3.
The sweep rate dependence is attributed to DNP occurring
in the breakdown regime of FQH states.
Results of a pump and probe experiment show that 
the polarities of the DNP induced in the breakdown regimes of the FQH states
is opposite to that of the DNP induced
in the breakdown regimes of odd-integer quantum Hall states.
\end{abstract}

\maketitle

Two-dimensional electron systems (2DESs)
subjected to perpendicular magnetic fields
exhibit the integer quantum Hall effect (QHE) 
with vanishing longitudinal resistance 
and quantized Hall resistance\cite{QHE}.
When a bias current applied to a 2DES exceeds a critical current ($I_{\rm c}$),
the quantum Hall conductor becomes unstable
against the excitation of electron-hole pairs,
and therefore,
the longitudinal resistance increases abruptly\cite{Ebert1983,Cage1983,Nachtwei1999,Komiyama2000}.
This phenomenon is referred to as the QHE breakdown.

In the case of breakdown of a quantum Hall (QH) state
with an odd-integer Landau level filling factor $\nu$,
electrons in the spin-up Landau subband
are excited to the spin-down subband, 
along with up-to-down flips of electron spins.
Since electron spins $\mathbf{S}$ interact with nuclear spins $\mathbf{I}$
via the hyperfine interaction
${\cal H}_{\rm hyperfine} = A\mathbf{S} \cdot \mathbf{I}
 = A(S^{+}I^{-} + S^{-}I^{+})/2 + AS_{z}I_{z}$,
where $A$ is the hyperfine constant,
electron spin flips cause nuclear spin flops.
Dynamic nuclear polarization (DNP)
in the breakdown of odd-integer QH states
has been demonstrated in our recent studies\cite{Kawamura2007,Takahashi2007}.
Nuclear spins are polarized
in the bulk part of the 2DES,
as shown by an experiment using a device with Corbino geometry\cite{Kawamura2008}.

Recent studies have revealed that fractional quantum Hall (FQH) states also
break down when a bias current exceeds $I_{\rm c}$\cite{Takamasu1995, Watts1998}.
Since the origin of the FQH effect, which is Coulomb interaction,
is different from that of the integer QHE, 
it is unclear whether DNP also occurs in the breakdown regimes of FQH states.
Furthermore, in FQH states,
competition between the exchange energy and  the Zeeman energy
leads to the production of a wide variety of ground states 
with different electron spin configurations\cite{Girvin}.
The different configurations
are expected to give rise to different DNP polarities.

In this paper, we report that DNP
occurs in the breakdown regimes of FQH states at $\nu$ = 2/3 and 1/3.
Results of our study indicate that
voltage-current characteristics
depend on  current sweep rates. 
The relationship between  DNP and the sweep rate dependence
is confirmed by nuclear magnetic resonance (NMR) measurements.
Furthermore, we determine the polarity and amplitude of the DNP
by a pump and probe experiment.
The DNP polarity is unexpectedly negative ($\langle I_z \rangle < 0$)
and is opposite to the polarity of DNP induced in the breakdown regimes
of odd-integer QH states\cite{Kawamura2007}.

\begin{figure}[b]
	\includegraphics[width=8.5cm]{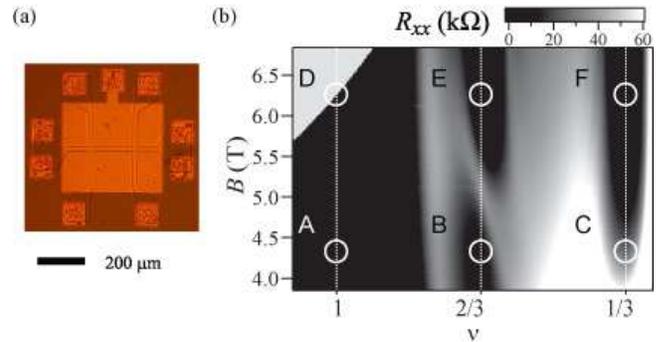}
	\caption{
	\label{concept}(Color online)
		(a) Micrograph of the Hall bar device used in the present study.
		(b) Gray-scale plot of the longitudinal resistance $R_{xx}$ 
			as a function of the external magnetic field $B$
			and the Landau level filling factor $\nu$ at 20 mK.
			$R_{xx}$ was measured by using a standard ac lock-in technique 
			at an excitation current of 
			$I_{\rm ac}$ = 1 nA (18 Hz). 
			The black regions represent quantum Hall states.
			Pump and probe measurements were performed
			at the conditions indicated by circles (A to F).
	}
\end{figure}

We used a GaAs/Al$_{0.3}$Ga$_{0.7}$As single heterostructure wafer
with a 2DES at the interface.
The mobility and sheet carrier density of the 2DES at 4.2 K
were $\mu$ = 228 m$^{2}$V$^{-1}$s$^{-1}$
and $n$ = 1.59 $\times$ 10$^{15}$ m$^{-2}$, respectively.
The wafer was processed into a 20-$\mu$m-wide Hall bar
covered with a front-gate electrode, as shown in  Fig.~\ref{concept}(a).
This electrode was used 
to change the Landau level filling factor of the 2DES.
Figure~\ref{concept}(b) shows a gray-scale plot of  the longitudinal 
resistance $R_{xx}$ as a function of the external magnetic field  $B$
and Landau level filling factor $\nu$.
The resistance peaks observed in the $\nu$ = 2/3 FQH state at around $B$ = 5.2 T
correspond to the spin transition, as reported
in previous studies\cite{Eisenstein, Engel}. 
The QH states on the low and high magnetic field sides of the transition
are spin-unpolarized and spin-polarized phases, respectively.

\begin{figure}[t]
	\includegraphics[width=7.0cm]{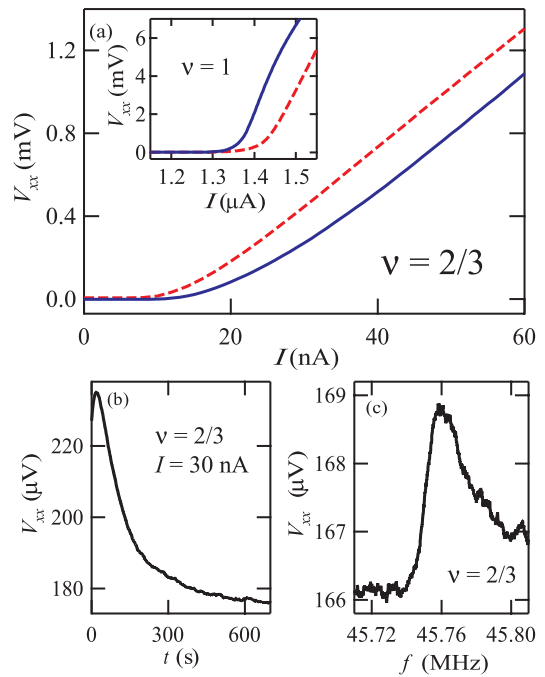}
	\caption{
	\label{cwNMR}(Color online)
		(a) $V_{xx}$-$I$  curves obtained 
		at $\nu$ = 2/3 and $B$ = 6.26 T [E in Fig.~\ref{concept}(b)].
		The current sweep rates were 6.8 nA/s (dashed) and 0.05 nA/s (solid).
		The inset shows $V_{xx}$-$I$ curves obtained at $\nu$ = 1 and $B$ = 6.26 T
		[D in Fig.~\ref{concept}(b)]
		with current sweep rates of 6.8 nA/s (dashed) and 0.4 nA/s (solid).
		(b) Time evolution of $V_{xx}$ after a sudden increase in $I$ 
		from 0 nA to 30 nA.
		(c) NMR spectrum of $^{75}$As detected by measuring $V_{xx}$ at
		 $\nu$ = 2/3 and $I$ = 30 nA.
 		}
\end{figure}

The dashed curve in the inset of Fig.~\ref{cwNMR}(a) represents 
a voltage-current ($V_{xx}$-$I$) characteristic curve
at $B$ = 6.26 T and $\nu$ = 1 [D in Fig.~\ref{concept}(b)].
The current was increased at a rate of 6.8 nA/s.
When the bias current $I$ exceeds a critical value
$I_{\rm c}$ = 1.4 $\mu$A, the longitudinal voltage $V_{xx}$ increases abruptly.
As mentioned earlier, this phenomenon is referred to 
as the QHE breakdown.
The $V_{xx}$-$I$ characteristic curve
represented by the solid curve in the inset of Fig.~\ref{cwNMR}(a)
was obtained by increasing the current with a sweep rate of 0.4 nA/s.
The QHE breaks down at a smaller $I_{\rm c}$ when the sweep rate is decreased.
The sweep rate dependence of the  $V_{xx}$-$I$ characteristic 
can be attributed to the DNP, as in the case of the
hysteretic $V_{xx}$-$I$ curves in our previous study\cite{Kawamura2007}:
In the breakdown regimes of odd-integer QH states,
electrons are excited to higher Landau levels,
and nuclear spins are dynamically polarized by the up-to-down flips of electron spins.
The polarized nuclear spins reduce
the spin-splitting energy $E_{\rm s} = |g^{*}|\mu_{\rm B}B - A\langle I_{z} \rangle$,
 where $g^{*}$ is the effective $g$-factor of electrons ($g^{*}$ = $-$0.44 in GaAs) 
and $\mu_{\rm B}$ is the Bohr magneton. 
Since the energy gap of odd-integer QH states is given by 
the sum of  $E_{\rm s}$ and exchange energy,
the reduction of $E_{\rm s}$ accelerates the QHE breakdown,
resulting in the shift of $V_{xx}$-$I$ curves toward the low current side\cite{Kawamura2007}.

Figure~\ref{cwNMR}(a) shows the $V_{xx}$-$I$ curves
obtained in the case of a spin-polarized FQH state
at $B$ = 6.26 T and $\nu$ = 2/3 [E in Fig.~\ref{concept}(b)].
The current sweep rates were 6.8 nA/s  and 0.05 nA/s 
for the dashed and solid curves, respectively.
As observed at $\nu$ = 1, $V_{xx}$ increases
when the current exceeds a critical value.
This indicates the breakdown of the FQH effect.
The FQH effect breaks down at a larger $I_{\rm c}$
when the sweep rate is decreased.
The sweep-rate-dependent $V_{xx}$-$I$ characteristics
appear similar to those observed in the breakdown regime of the $\nu$ = 1 QH state,
except for the direction of the shift of the $V_{xx}$-$I$ curves.
Figure~\ref{cwNMR}(b) shows the time evolution of $V_{xx}$
at the $\nu$ = 2/3 spin-polarized FQH state after a sudden increase in $I$ 
from 0 nA to 30 nA.
The value of $V_{xx}$ decreases slowly over a period of 600 s,
which is a typical time scale 
for nuclear spin related phenomena\cite{Kawamura2007,Song2000,Machida2003,Kronmuller1998}.
Thus, the slow evolution of $V_{xx}$
and the sweep-rate-dependent $V_{xx}$-$I$ curves
indicate that  DNP occurs in the breakdown regime of the FQH state.

The relationship between the DNP
and the sweep-rate-dependent $V_{xx}$-$I$ characteristics 
and slow evolution of $V_{xx}$ was determined by NMR measurements as follows.
In order to apply a radio-frequency (rf) magnetic field $B_{\rm rf}$ 
perpendicular to the external magnetic field $B$ (parallel to the 2DES),
we wound a single-turn coil around the Hall bar device.
As the frequency of $B_{\rm rf}$ was scanned, 
$V_{xx}$ peaked at the NMR frequencies of $^{75}$As, 
as shown in Fig.~\ref{cwNMR}(c).
NMR spectra of $^{69}$Ga and $^{71}$Ga were also obtained.
Similar sweep-rate dependence in $V_{xx}$-$I$ characteristics 
and NMR were observed
at all the QH states investigated [A to F in Fig.~\ref{concept}(b)].
The detection of NMR by the voltage measurements definitely 
revealed the occurrence of DNP in the breakdown regimes of the FQH states.

It should be noted that the DNP observed in this study is different from
the well-known DNP\cite{Kronmuller1998,Kronmuller1999,Kraus2002,
Stern2004,Hashimoto2002,Hashimoto2004}
occurring near the spin transition at $\nu$ = 2/3.
It has been suggested that
the DNP near the spin transition is induced 
by electron scattering between spatially distributed spin-polarized
and spin-unpolarized domains\cite{Kronmuller1999, Kraus2002}.
Therefore, the coexistence of spin-polarized and spin-unpolarized domains
is considered a prerequisite for the DNP near the spin transition.
In contrast, the DNP observed in this study occurs
in the breakdown regime of the FQH states
 away from the spin transition point.
In the condition, spin configuration of the FQH system
is not affected by the spin transition
and the complex spin-domain structure does not exist.
We think that a mechanism other 
than the inter-spin-domain scattering\cite{Kronmuller1999, Kraus2002}
is needed to understand the DNP observed in this study.

As shown in Fig.~\ref{cwNMR}(a),
the $V_{xx}$-$I$ curves shift as the DNP occurs.
The directions of the shifts at $\nu$ = 2/3 and $\nu$ = 1 are opposite to each other;
this indicates that the DNP polarities are different.
Therefore, in order to investigate the polarities of DNP more systematically,
we conducted the pump and probe experiment:
First, the 2DES was temporarily set to a breakdown state ($\nu_{\rm pump}, I_{\rm pump}$)
for a duration $\tau_{\rm pump}$ in order to induce DNP.
Next, the bias current was turned off, and the filling factor was suddenly changed 
to $\nu_{\rm probe} = 1$ by using the front-gate electrode.
Then, the $I_{\rm c}$ value of the $\nu_{\rm probe} = 1$ QH state
was measured by increasing the bias current\cite{procedure}.
Since the $I_{\rm c}$ value is expected to change in accordance with $\langle I_z \rangle$,
we can determine the polarity and  amplitude of DNP induced at $\nu_{\rm pump}$\cite{procedure2}.

\begin{figure}[t]
	\includegraphics[width=8.5cm]{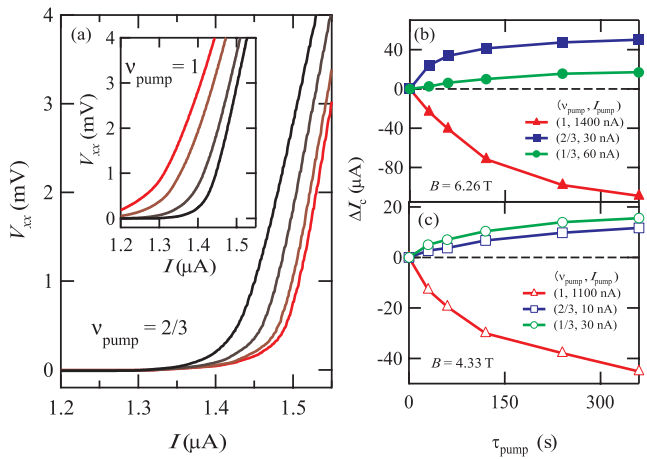}
	\caption{
	\label{PIV}
		(Color online)
		(a)	$V_{xx}$-$I$ curves obtained at $\nu$ = 1 and $B$ = 6.26 T
		after inducing DNP at ($\nu_{\rm pump}$, $I_{\rm pump}$) = (2/3, 30 nA)
		for $\tau_{\rm pump}$ values 
		of 0, 30, 120, and 360 s (from left to right).
		Inset: $V_{xx}$-$I$ curves obtained at $\nu$ = 1 
		after inducing DNP at ($\nu_{\rm pump}$, $I_{\rm pump}$) = (1, 1400 nA).
		The values of $\tau_{\rm pump}$ are 0, 30, 120, and 360 s from right to left.
		(b)
		Horizontal shift $\Delta I_{\rm c}$ of $V_{xx}$-$I$ curves defined at $V_{xx}$ = 2 mV
		plotted as a function of $\tau_{\rm pump}$ at $B$ =  6.26 T.
		Results obtained at ($\nu_{\rm pump}$, $I_{\rm pump}$) =
		(1, 1400 nA), (2/3, 30 nA), and (1/3, 60 nA) 
		are represented by solid triangles, solid squares, and solid circles, respectively.
		(c)
		Similar data obtained at $B$ = 4.33 T.
		Results obtained at  ($\nu_{\rm pump}$, $I_{\rm pump}$) = (1, 1100 nA),
		(2/3, 10 nA), and (1/3, 30 nA)
		are represented by open triangles, open squares, and open circles, respectively.
	}
\end{figure}

The inset of Fig.~\ref{PIV}(a) shows $V_{xx}$-$I$ curves obtained at $\nu$ = 1
after inducing DNP at ($\nu_{\rm pump}$,  $I_{\rm pump}$) = (1, 1400~nA) and $B$ = 6.26~T
[D in  Fig.~\ref{concept}(b)] for several values of $\tau_{\rm pump}$.
The rightmost curve (black) represents the results obtained without inducing DNP
($\tau_{\rm pump}$ = 0~s),
and the leftmost curve (red) represents the results obtained  after inducing DNP 
for $\tau_{\rm pump}$ = 360~s.
The horizontal shift $\Delta I_{\rm c}$ of the $V_{xx}$-$I$ curve defined at $V_{xx}$ = 2~mV
is plotted as a function of  $\tau_{\rm pump}$ [closed triangles in Fig.~\ref{PIV}(b)].
$\Delta I_{\rm c}$  decreases with increasing $\tau_{\rm pump}$.
This indicates that the decrease in $\Delta I_{\rm c}$
can be attributed to the DNP.

The main panel of Fig.~\ref{PIV}(a) shows $V_{xx}$-$I$ curves obtained at $\nu$~=~1
after inducing DNP  at ($\nu_{\rm pump}$, $I_{\rm pump}$) = (2/3, 30~nA) and $B$~=~6.26~T
[E in Fig.~\ref{concept}(b)]
for several values of  $\tau_{\rm pump}$.
The curves shift to higher currents with increasing $\tau_{\rm pump}$.
The $\tau_{\rm pump}$ dependence of  $\Delta I_{\rm c}$
is indicated by solid squares in Fig.~\ref{PIV}(b).  
$\Delta I_{\rm c}$ increases with  $\tau_{\rm pump}$. 
It should be noted that $\Delta I_{\rm c}$ decreases when an rf magnetic field with
NMR frequency is applied.
Therefore, the nonzero values of $\Delta I_{\rm c}$ indicate that
DNP occurs in the breakdown regimes of the FQH states\cite{heating}.
Data obtained on the spin-unpolarized phase at $B$ = 4.33 T [B in Fig.~\ref{concept}(b)]
are shown by open squares in Fig.~\ref{PIV}(c).
The sign of $\Delta I_{\rm c}$ is also positive for $B$ = 4.33 T.
Similar data obtained for $\nu_{\rm pump}$ = 1/3 
at $B$ = 6.26 T [F in Fig.~\ref{concept}(b)]
are shown by solid circles in Fig.~\ref{PIV}(b).
In this case also, $\Delta I_{\rm c}$ increases with  $\tau_{\rm pump}$.

\begin{figure}[t]
	\includegraphics[width=8.5cm]{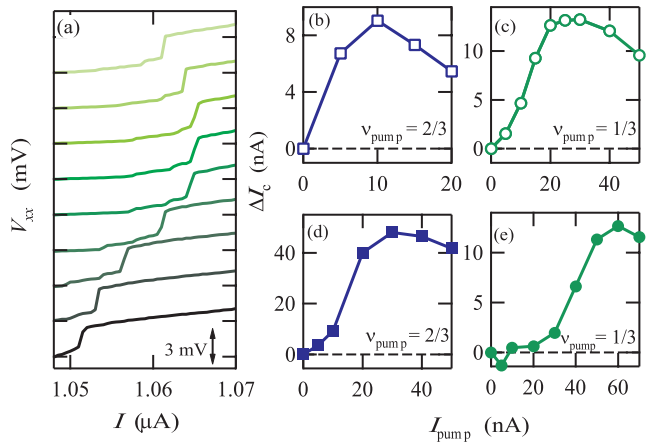}
	\caption{
	\label{ipump}
	(a) $V_{xx}$-$I$ curves obtained after inducing DNP at
	($\nu_{\rm pump}$, $B$) = (1/3, 4.33 T) and $\tau_{\rm pump}$ = 600 s
	with different pump currents:
	$I_{\rm pump}$ = 0, 5, 10, 15, 20, 25, 30, 40, and 50 nA
	(from bottom to top). The curves are offset vertically for clarity.
	(b)--(e) Horizontal shift $\Delta I_{\rm c}$ of $V_{xx}$-$I$ curves plotted
	as a function of $I_{\rm pump}$
	at ($\nu_{\rm pump}$, $B$) = (2/3, 4.33 T), (1/3, 4.33 T), (2/3, 6.26 T), 
	and (1/3, 6.26 T), respectively.
	}
\end{figure}

Figure~\ref{ipump}(a) shows $V_{xx}$-$I$ curves obtained after inducing DNP at
($\nu_{\rm pump}$, $B$) = (1/3, 4.33 T) and several values of $I_{\rm pump}$ 
for $\tau_{\rm pump}$ = 600 s.
The $V_{xx}$-$I$ curves shift horizontally with increasing $I_{\rm pump}$.
The values of $\Delta I_{\rm c}$ 
are plotted as a function of $I_{\rm pump}$ in Fig.~\ref{ipump}(c) (open circles).
The horizontal shift  $\Delta I_{\rm c}$ is maximum at $I_{\rm pump}$ = 30 nA.
Qualitatively similar data are obtained for
($\nu_{\rm pump}$, $B$) = (2/3, 4.33 T), (2/3, 6.26 T), and (1/3, 6.26 T),
as shown in Figs.~\ref{ipump}(b), (d), and (e), respectively.
In all cases, the amplitude of DNP increases steeply when $I_{\rm pump}$ exceeds $I_{\rm c}$
and is maximum when $I_{\rm pump}$ is slightly greater than $I_{\rm c}$.
The $I_{\rm pump}$ dependence of DNP in the breakdown regime of FQH states 
is similar to that observed in the breakdown regime of odd-integer QH states\cite{Kawamura2008}.
This similarity indicates that the origin of DNP occurring in the breakdown regimes of both FQH and odd-integer QH states is the same.

We discuss the polarities 
of the DNPs occurring in the breakdown regimes of the FQH states.
The positive signs of the $\Delta I_{\rm c}$ values
for $\nu_{\rm pump}$ = 2/3 and 1/3 are opposite to those  for $\nu_{\rm pump}$ = 1.
The positive sign of $\Delta I_{\rm c}$ value shows that
the energy gap of the $\nu_{\rm probe}$ = 1 QH state increases due to DNP.
Since $E_{\rm s}$ = $|g^{*}|\mu_{\rm B}B - A\langle I_z \rangle$
increases due to negative DNP ($\langle I_z \rangle < 0$),
DNP with negative polarity is deduced from the positive $\Delta I_{\rm c}$ value.
The results in Figs.~\ref{PIV}(b) and ~\ref{PIV}(c)
show that DNPs with negative polarities are induced in the breakdown regimes 
of $\nu_{\rm pump}$ = 1/3 and $\nu_{\rm pump}$ = 2/3.

The negative DNPs  for $\nu_{\rm pump}$ = 1/3,
opposite to the DNPs for $\nu_{\rm pump}$ = 1, are unexpected
because electron spins are fully polarized in both ground states.
Furthermore, for $\nu_{\rm pump}$ = 2/3, the signs of DNPs are the same (negative)  both 
on the spin-unpolarized phase at 4.33 T 
and on the spin-unpolarized phase at 6.26 T [B and E in Fig.~\ref{concept}(b), respectively],
despite of the difference in electron spin configurations in the ground states.
These results suggest that neither electron spin configurations 
in the FQH ground  states nor spin transitions are relevant 
to the polarity of DNP occurring in the breakdown regimes.
In the breakdown regime, a number of electron-hole pairs are excited in the 2DES
and the electron spin configurations of the FQH ground states,
which is based on subtle electron-electron correlation,
are probably not maintained.
We infer that spin dynamics of these excitations
plays an important role to understand the mechanism of the DNPs occurring 
in the breakdown regimes.

This study is supported by a Grant-in-Aid from 
MEXT, the Sumitomo Foundation, 
and the Special Coordination Funds for Promoting 
Science and Technology.


\begin{thebibliography}{99}
\bibitem{procedure}
	The current was increased rapidly
	so that DNP during the current sweep was negligible.
\bibitem{procedure2}
	DNPs induced at various $\nu_{\rm pump}$ were detected 
	at the fixed $\nu_{\rm probe}$ = 1 so that we can discuss 
	the relative polarities and amplitudes of the DNPs.
\end{thebibliography}
\end{document}